\documentclass[preprint]{aastex} 

\begin{document}

\title{IS THE BLAZHKO EFFECT THE BEATING OF A NEAR-RESONANT DOUBLE-MODE PULSATION?}

\author{Paul H. Bryant}
\affil{BioCircuits Institute (formerly Institute for Nonlinear Science), \\University of California, San Diego, La Jolla, CA 92093, USA}
\email{pbryant@ucsd.edu}

\begin{abstract}
In this paper it is shown that the Blazhko effect may result from a near-resonant type of multi-mode pulsation, where two (or sometimes more) periodic oscillations with slightly different frequencies gradually slip in phase, producing a beat frequency type of modulation.  Typically one of these oscillations is strongly non-sinusoidal.  Two oscillations are sufficient for the standard Blazhko effect; additional oscillations are needed to explain multi-frequency modulation.  Previous work on this hypothesis by Arthur N. Cox and others is extended in this paper by developing a simple (non-hydro) model that can accurately reproduce several important features found in \emph{Kepler} data for RR Lyr, including the pulsation waveform, the upper and lower Blazhko envelope functions and the motion, disappearance and reappearance of the bump feature.  The non-sinusoidal oscillation is probably generated by the fundamental mode and the other oscillations are probably generated by nonradial modes.  This model provides an explanation for the strong asymmetry observed in the side peak spectra of most RR Lyrae stars.  The motion and disappearance of the bump feature are shown to be an illusion, just an artifact of combining the oscillations.  V445 Lyr is presented as an example with dual modulation.  The mysterious double-maxima waveform observed for this star is explained, providing additional strong evidence that this beating-modes hypothesis is correct.  Problems with other recent explanations of the Blazhko effect are discussed in some detail.
\end{abstract}


\section{INTRODUCTION}
\label{introduction}
Variable stars of the RR Lyrae type often exhibit a slow modulation of their pulsations known as the Blazhko effect.  First reported in 1907 by Russian astronomer Sergey Nikolaevich Blazhko \citep{Blazhko}, the effect remains mysterious, with none of the numerous proposed explanations achieving widespread acceptance.  See the recent reviews by \citet{Kovacs} and \citet{Kolenberg2012}.  Recent attempts to explain the Blazhko effect include:
\begin{itemize}
\item ``The ninth overtone model" which involves an unstable resonance interaction between between the fundamental mode and the ninth radial overtone, the oscillations of which lead to modulation of the fundamental \citep{Szabo, Buchler2011, Kollath2011}
\item ``The Stothers model" in which an oscillatory magnetic field induces variations in the strength of turbulent convection which in turn modulates the fundamental \citep{Stothers2006, Stothers2010}
\item ``The shock model" in which a shock wave periodically generates turbulence in the atmosphere which in turn modulates the fundamental \citep{Gillet}.
\end{itemize}
All of these models have significant problems which are discussed in Section~\ref{problems}.

\begin{figure}
\begin{center}
\includegraphics[width=6.0in]{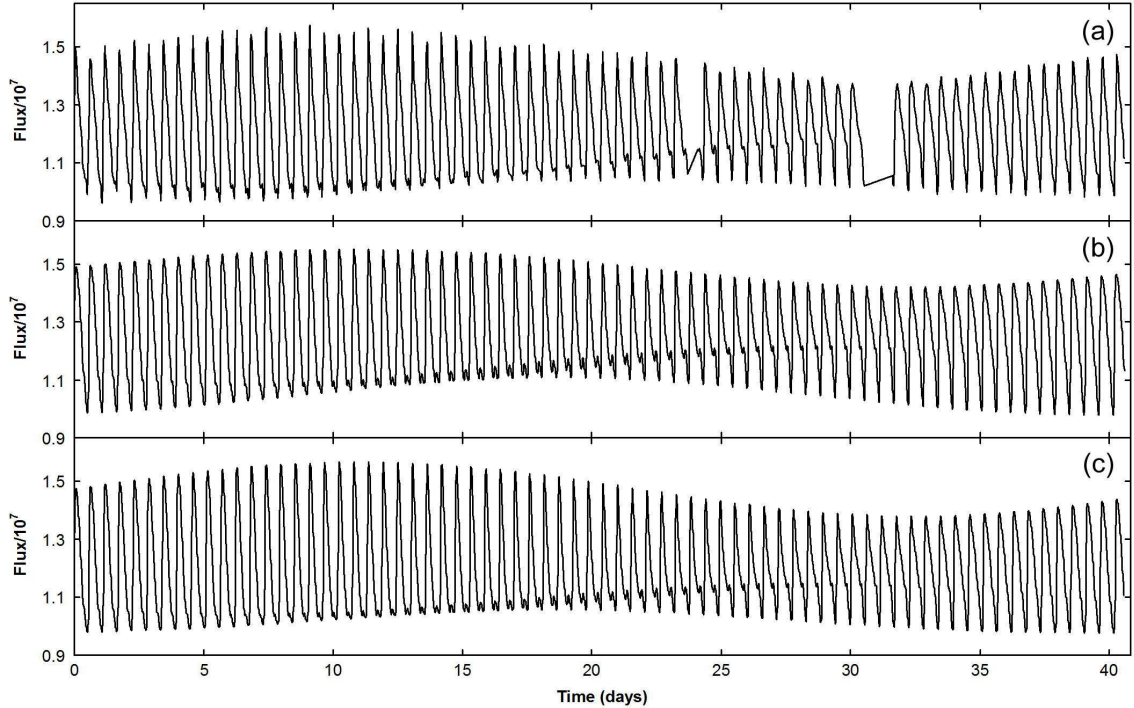}
\caption{\label{dam}Comparison of data with the model. (a) Data from the \emph{Kepler} project for RR Lyr (KIC 7198959, second quarter, long cadence corrected flux data starting at TJD 55032.7625503).  72 pulsation cycles are displayed which constitutes approximately 1 Blazhko cycle.  There are two visible gaps where good data is not available.  (b) Model output generated by adding 2 periodic oscillations, PO1 and PO2.  PO1 is a highly non-sinusoidal waveform, generated as discussed in the text and shown in Figure~\ref{po1}.  PO2 is a simple sine wave of slightly higher frequency than PO1.  (c) Corrected model as discussed in the text.  Note that the model and corrected model match many features of the data including the phase offset of the upper and lower Blazhko envelope curves (with the lower one leading), the shape of the pulsation waveform, the apparent motion of the bump feature from the bottom upwards towards the middle of the waveform, its subsequent disappearance and then reappearance at the bottom.  The corrected model also matches the dissimilar amplitudes of the upper and lower Blazhko envelopes.}
\end{center}
\end{figure}

In this paper a simple explanation is put forth for the Blazhko effect which involves (at least) two periodic oscillations, PO1 and PO2, of slightly different frequency.  Typically one of these oscillations is strongly non-sinusoidal.  The slow phase slip between these oscillations results in a combined waveform that displays a beat frequency or apparent modulation at the Blazhko frequency.  We refer to this as ``the beating-modes model".  It is the only model where there is no actual modulation of the amplitude of the primary mode, rather \emph{the apparent modulation is an illusion} produced by the partial cancellation that results from adding to it the other mode (or modes) which have nearly the same frequency but differing phases. This hypothesis is supported by a simple (non-hydro) model of the dynamics that accurately reproduces \emph{Kepler} data for RR Lyr, as shown in Figure~\ref{dam}\notetoeditor{Figure~\ref{dam} is intended to be full page width, not single column}.  In addition, the beating-modes model provides an explanation for the mysterious ``double-maxima" waveform of V445 Lyr (See Section~\ref{multifreq}), providing additional strong evidence in its favor.

Referring to Figure~\ref{dam}, note the many features of the data that are reproduced by the model including the shape of the pulsation waveform, the phase offset between the upper and lower modulation envelope functions, the size of the bump feature, the behavior of the bump: hesitating near the bottom of the pulsation waveform, moving upwards for a while, fading out and disappearing for a while and then reappearing near the bottom to complete the cycle.  The disappearance of the bump lasts for about a third of the Blazhko cycle starting near the Blazhko minimum.  The bump behavior of the model is a little surprising since it is entirely brought about by the addition of PO2, a pure sinusoid.  PO1 by itself displays a bump at a fixed location about one quarter up from the bottom.

PO1 and PO2 may be generated by two unstable modes of the star.  It is generally believed that the pulsation frequency for Blazhko stars corresponds to the fundamental mode, so this would be the likely choice for PO1, while PO2 may be generated by a nonradial mode of nearly the same frequency.  An alternate possibility is that PO2 is a stable nonradial mode which becomes unstable by acquiring additional energy from the fundamental.  (This is only possible if a mechanism exists that can accomplish this without phase-locking.)  Since nonradial modes are involved, the usual 1D radial hydrocode software will not be able to test this model.  Multi-D codes are now being developed which may be able to accomplish this at some future date \citep[see e.g.][]{Mundprecht2013,Geroux2014}.

There have been several previous papers in support of the beating-modes model, but the current work goes much further in terms of finding a simple model that actually provides a remarkably good fit to the observed Blazhko data, including the mysterious motion and disappearance of the so-called bump feature.  \citet{Guggenberger2008}, building on earlier work by \citet{Breger2006}, proposed a test of the phasing behavior in an effort to determine if beating was responsible for the Blazhko effect.  They applied this to the two well-observed Blazhko RR Lyrae stars SS For and RR Lyr, with results consistent with the beating of two oscillations.  \citet{Cox1993}, based on a combination of nonradial modal stability analysis combined with radial hydrocode analysis, proposed a near-resonant beating process between the fundamental and a nonradial mode, most likely the $g_4$, $l=1$ mode.  (Note that the stability analysis was for a static star and so might be significantly impacted by the presence of the high amplitude fundamental.)  Sixteen years later, \citet{Cox2009}, working with improved code, finds that nonradial modes near the fundamental are always just slightly stable, and proposes that the $g_4$, $l=1$ mode becomes unstable through nonlinear interaction with the fundamental.  He still refers to this as a beating process not as a resonant mode interaction.  \citet{Borkowski} proposed a near-resonant double-mode pulsation involving the fundamental and either the second or third radial overtone, where the overtone frequency is higher than the second harmonic of the fundamental by the Blazhko frequency.  Because the overtone is near the second harmonic instead of the fundamental itself, the Borkowski model is not validated by the results of this paper.  Also since it only involves radial modes, one might have expected it to have been observed in 1D hydrocode models, but apparently it has not.

A variant of the beating-modes model is found in the work of \citet{Nowakowski}.  Their approach involves a resonance between the fundamental mode and a pair of nonradial modes with opposite orders $\pm m$.  Modes with nonzero $m$ have oscillations as a function of the azimuthal angle.  These can be thought of as traveling wave modes that move around the star either with or against the rotational motion of the star depending on the sign of $m$.   As a result, these are split in frequency by the rotation.  These modes can derive energy through nonlinear interaction with the fundamental if they satisfy the resonance condition that the sum of their frequencies must be exactly twice the fundamental frequency.  As a result they generate a triplet spectrum with one peak on either side of the fundamental.  According to Equation 77 of \cite{Nowakowski}, the resulting Blazhko frequency is some fraction of the rotational frequency, most likely around 1/2.  The pair are assumed initially stable and made unstable through the resonance.  The resonant interaction between the modes is presumed stable and therefore the mode amplitudes and frequencies do not vary in time.  Since all three peaks of the triplet correspond to actual modes, this is another type of beating-modes model, where the combined waveform appears modulated even though no actual modulation taking place.  The main problem with this model is that there seems to be no good explanation for the amplitude disparity that is often observed for the opposite side peaks.  However it seems reasonable to consider, due to the wide variation of behavior observed among Blazhko stars, that this model could apply to some stars which have a relatively balanced triplet.

The spectra for Blazhko stars includes progressions of side-peaks to the left and right of the fundamental and each of its harmonics, all evenly spaced apart by the Blazhko frequency.  These tend to diminish in amplitude when moving away to the left or right and often only the first side-peak is visible above the noise on each side.  Most side-peaks are nonlinear mixing product peaks and do not correspond to actual excited modes of the star.  Usually these side-peaks are attributed to modulation of the fundamental mode.  However, if at least one of the side-peaks immediately adjacent to the fundamental corresponds to an actual mode, this is sufficient through nonlinear processes (as discussed below) to generate the entire set of side-peaks.  The other frequencies are integer linear combinations of these two frequencies:
\begin{equation}\label{side-peak}
\omega_{nm} = n \omega_1 + m \omega_2,
\end{equation}
where $\omega_{nm}$ is a side-peak frequency, $n$ and $m$ are integers (positive or negative), $\omega_1$ is the frequency of PO1 (the fundamental) and $\omega_2$ is the frequency of PO2.  So, for example, the side-peak that is opposite to $\omega_2$ (on the other side of the fundamental) has $n=2$ and $m= -1$.
  If one side-peak is notably larger than the others then it is the likely choice for PO2.  In the case of RR Lyr, there is a significantly larger peak on the higher frequency side.
A spectral analysis (using the Period04 software package) of the Kepler data used in Figure~\ref{dam}(a) gives the amplitude (in units of $10^7$) of the upper peak as 0.0487 and of the lower peak as 0.0097, giving an amplitude ratio of about 5.02 and corresponding power ratio of 25.2.  The central peak amplitude is about 0.1777.  

It is well known that amplitude and frequency modulation can generate a side-peak spectrum similar to the one observed \citep[see e.g.][]{Benko2011,Szeidl2009,Szeidl2012}, however a quasiperiodic dynamical system will also generate such a spectrum.  A quasiperiodic system is one whose oscillations contain two base frequencies that are not rationally related.  These would correspond to the frequencies of PO1 and PO2 and the apparent modulation frequency is the difference between them.  The complete spectrum of such a system consists of the set of frequencies given by Equation~\ref{side-peak}.  It can be obtained by a two-dimensional Fourier series expansion as is shown in Appendix~\ref{quasi}.  Nonlinear processes can change the amplitudes of the (side-peak) frequencies including those which are initially zero.  We take as a starting point a spectrum that includes PO1 (the fundamental mode frequency with its harmonics) and PO2 (a sine wave).  There are two nonlinear mechanisms for generating the complete side-peak array.  One is the process that translates the pulsational motion of the star into a variation in the light output.  There is no reason to assume this process is entirely linear.  So, for example, if the light output depends slightly on the square of the sum of the radial velocities of the two oscillations, then there will be cross terms generated by the square which will introduce components at the triplet (or first order) side-peaks for the fundamental and all of its harmonics.  Higher order nonlinear terms will generate higher order side-peaks.  A second mechanism is generated by nonlinear coupling between the active modes.  Although such coupling is (by assumption) insufficient to produce phase-locking between the modes, it can introduce side-peaks that were previously missing, change the amplitudes of peaks that were already present and shift the base frequencies.  In the first mechanism, the side-peak frequencies are only present in the light curve, while in the second they are also physically present in the stellar vibration.  Both mechanisms may be active at the same time.

\section{MODEL}
\label{model}
For purposes of matching the \emph{Kepler data, }PO1 and PO2 are approximated as constant in amplitude and frequency throughout the Blazhko cycle, which is equivalent to saying that the nonlinear interactions between them are neglected as relatively small.  PO1 is a highly non-sinusoidal waveform whose frequency $\omega_1$ is the pulsation frequency.  Based on success at fitting the data with the model, it appears that this oscillation is made up of two sub-components which we will call PO1a and PO1b both of which are sawtooth-like waveforms.  PO1a is larger in amplitude, while PO1b is shifted in phase and generates the bump feature. The most likely possibility is that PO1 is generated by a single mode with complicated dynamics.  It is generally thought that the bump is an echo, either a reflection from the stellar core or the atmosphere \citep{Guggenberger2006} and thus PO1b is the echo of PO1a.   An alternate, but less likely, possibility is that PO1b is actually a separate mode, probably nonradial, that is phase-locked to the fundamental.  This would make it similar to Cepheids which are thought to exhibit a bump due to a resonance with another mode \citep{Simon,Buchler1990,Bono2000}.  But regardless of the reason for the bump, the important point is that the model generates a reasonably good approximation to the observed waveform, enabling a test of the beating-modes hypothesis.  The model output for PO1 is shown in Figure~\ref{po1}.  Details of how it is generated are given below.
\begin{figure}
\begin{center}
\includegraphics[width=1.0\columnwidth]{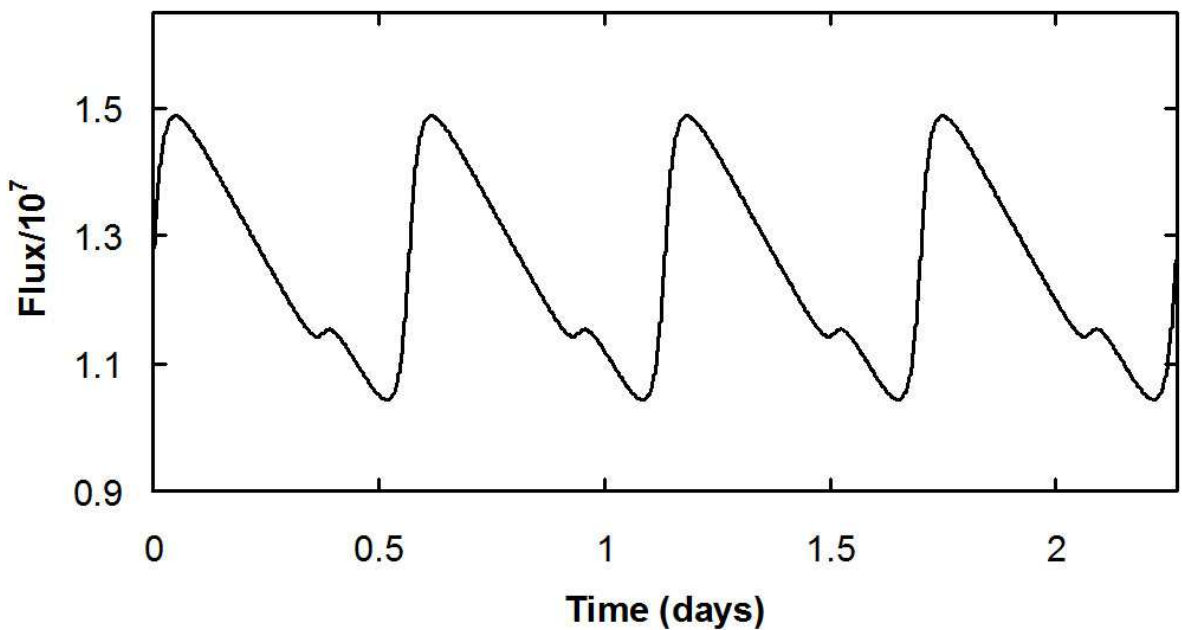}
\caption{\label{po1}Shows the first periodic oscillation PO1, offset by the RR Lyr base flux level of $1.25 \times 10^7$.  When this is added to PO2, a simple sine wave, it produces the model output shown in Figure~\ref{dam}(b).  PO2 is slightly higher in frequency than PO1, the difference being the Blazhko frequency.  Note that the bump feature is stationary in PO1.  The apparent motion and disappearance of the bump in Figure~\ref{dam} is simply an artifact produced by the addition of PO2.}
\end{center}
\end{figure}
PO2, on the other hand, is approximated as a pure sinusoid as this seems to do a good job of reproducing the observed Blazhko behavior.  The frequency of PO2 is $\omega_2=\omega_1 + \omega_B$, where $\omega_B$ is the Blazhko frequency.  In Section~\ref{relation} it is discussed how the current work relates to the ``hybrid mode model" developed previously by the author \citep{Bryant}.

For PO1a and PO1b we will use the same one-zone model developed in \citet{Bryant}, although here we are only using the model to generate the sawtooth-like waveform needed for PO1.  The model details are given in Appendix~\ref{one-zone}.  This one-zone stellar model is similar to one developed by \cite{Baker} and used extensively by Stellingwerf and others \citep{Stellingwerf1972, Stellingwerf1986, Stellingwerf1987, Munteanu} to study Cepheid and RR Lyrae variable stars.  When excited to high amplitude, this model produces a sawtooth-like velocity function, a waveform that is commonly seen in the light curves of RR Lyrae and Cepheid variable stars.  For simplicity, we will use identical model parameters to generate PO1a and PO1b, the only differences are that PO1b has its amplitude reduced by a scale factor and it is shifted in phase relative to PO1a.  Both sub-components are made equally non-sinusoidal, which seems to produce a very good fit to the observational data.

Let $v_c$ be the combined velocity of PO1 and PO2.  For PO1 we use $v_1(t)$ from Equation~\ref{eqnpo1} in Appendix~\ref{one-zone} and for PO2 we use a sine wave, with phase and scale factor chosen to fit the \emph{Kepler} data:
\begin{equation}\label{combined}
v_c(t) =  v_1(t) +0.05sin(\omega_2 t -0.6)
\end{equation}
The light flux output $L(t)$, as plotted in Figure~\ref{dam}(b), is obtained from:
\begin{equation}\label{light}
L(t)/10^7=1.25+1.3v_c(t).
\end{equation}
The coefficients used were selected to obtain good visual agreement with the \emph{Kepler} data in Figure~\ref{dam}(a)

We observe that for the \emph{Kepler} data, the upper and lower Blazhko envelopes have differing amplitudes with the lower one being considerably smaller than the upper.  The model on the other hand has upper and lower amplitudes of the envelope functions that are equal to each other and to the amplitude of PO2.  We take a phenomenological approach to this problem and assume that the observed stellar flux is approximately equal to some nonlinear function of the combined radial velocity.  In particular we add a correction term proportional to $v_c^2(t)$ to the expression for the light flux.  The result is shown in Figure~\ref{dam}(c) and obtained from
\begin{equation}\label{lightcorr}
L_{corr}(t)/10^7=1.18+1.3v_c(t)+1.56v_c^2(t)
\end{equation}
The effect of the correction term is to stretch the upper part of the graph increasing the amplitude of the upper envelope and compress the lower part decreasing the amplitude of the lower envelope. Note that there may be other ways to achieve the same effect, e.g. it could conceivably result from a nonlinear interaction between the modes that generate PO1 and PO2.

\section{OTHER BLAZHKO CASES}
\subsection{Multiple Frequency Blazhko}
\label{multifreq}
Recent work \citep{Skarka,Benko} has shown that it is fairly common to find Blazhko cases that have two or more simultaneous modulation frequencies.  This is accompanied with the appearance in the spectra of two or more sets of side-peaks with different spacings.  This could be the result of having an additional excited mode or modes near the fundamental that is not phase-locked to one of the peaks in the original side-peak array.  Nonlinear processes can then generate additional side peaks that are integer linear combinations of these with the other base frequencies.

A good example of a complex case is V445 Lyr \citep{Guggenberger2012}, which exhibits a dual frequency Blazhko modulation.   The spectrum shows two strong peaks just above the fundamental and much weaker peaks below it.  Thus it is likely that these strong peaks correspond to two active modes in addition to the fundamental.  Each mode has an associated set of side-peaks.  Modeling in this case would require the addition of all three modal oscillations, which could be labeled PO1, PO2 and PO3.  With this simple change, the model will generate the observed dual frequency Blazhko modulation.  When the amplitude is large, the observed waveform has the typical sawtooth-like character \citep[see e.g. Figure~1(b) in][]{Guggenberger2012} although its rising section is not nearly as sharp as in the waveform for RR Lyr.

Occasionally the Blazhko minimum of V445 Lyr is very low, at which point it exhibits a ``double-maxima" waveform, see \citep[see e.g. Figures~1(b) and 1(d) in][]{Guggenberger2012}.  This can be easily explained as a cancellation of the base frequency, leaving the harmonic content behind.  The first oscillation, PO1 is strongly non-sinusoidal and hence has a significant harmonic content.  It can be thought of as the sum of two components: a sine wave at the base frequency and the harmonic content (2nd harmonic and above).  The other two oscillations PO2 and PO3 have little harmonic content and so can be approximated as sine waves with frequencies that are very close to the base frequency of PO1.  If the sum of the amplitudes of PO2 and PO3 is approximately equal to the amplitude of the base frequency component of PO1, then approximate cancellation is possible.  It will occur when PO2 and PO3 are approximately in phase with each other and approximately 180 degrees out of phase with the base component of PO1.  Since the relative phases of PO1, PO2 and PO3 are slowly changing, the near cancellation will only persist for a short time interval (a few pulsation cycles). The harmonic content of PO1 has nothing to cancel with and therefore remains, producing for that short interval a waveform that appears to be almost entirely harmonics.  This is the observed double-maxima waveform.  The doubling of the maximum suggests a strong second harmonic content in PO1.

Explanations such as the ninth overtone model that involve an actual modulation of the fundamental would appear to be unable explain this effect.  Normally when a nonlinear oscillation is at a low amplitude the nonlinear effects become dramatically less important.  In the limit of small amplitude, the oscillation is expected to approach a perfect sine wave.  But here the opposite is apparently happening---the amplitude is very small, but the harmonics, and particularly the second harmonic, are stronger than the base frequency.

Assuming for the sake of argument that the ninth overtone is a viable model (in spite of the serious problems outlined in Section~\ref{ninthmodel}), one might try to explain the strong harmonic content as resulting from interactions between the fundamental and the ninth overtone modes.  But these interactions are nonlinear and in the small amplitude limit all such interactions vanish and the vibrational modes become completely independent.  In this regard we note that the double-maxima waveform appears only when the apparent amplitude of the fundamental mode is at it smallest, probably about 10 times smaller that at its maximum.  We also note that the peak in the observed spectrum that (supposedly) corresponds to the ninth overtone mode is extremely small in amplitude, see $f_H + 3f_0$ in Table 2  in \citet{Guggenberger2012}.  Thus, within the ninth overtone model, it does not seem reasonable that sufficient nonlinear distortion can be present to account for the double-maxima waveform.

\subsection{Different Triplet Symmetry}
A study of variable stars in the Large Magellanic Cloud \citep{Alcock} found 74\% of Blazhko stars had a larger side-peak on the higher frequency side.  These cases are potentially compatible with the RR Lyr results, but what about the other 26\%?  Since nothing presented here depends on the active side-peak being on the high side, one possibility that the active mode is on the lower frequency side.

It is also possible that some stars may have additional side-peaks that are active modes.  If another mode is phase-locked to one of the side-peaks (or to one of the harmonics of the fundamental) it may be difficult to determine with certainty that it is present.  One indication of this could be that the side-peak in question is notably higher in amplitude than expected from the pattern of its neighbors.

In cases where the side peaks are nearly symmetric in amplitude, the model of \cite{Nowakowski} could be at work.  This is a beating-modes model that is generated by a three mode resonance.  See the discussion of this model in Section \ref{introduction}.

Cases of extremely low amplitude Blazhko modulation could potentially be generated by turbulence noise.  This could be the case when there is a stable resonance between the fundamental and a nonradial mode at the same frequency.  These modes would normally be locked at a particular relative phase.  However, if the phase is offset from equilibrium it may oscillate back and forth as it decays back to the equilibrium point.  Such oscillations are typically seen in amplitude equation simulations, see e.g. the introduction of \citet{Buchler1984}.  So there is phase modulation, but when adding the two modes together a beating process will occur generating the appearance of amplitude modulation.  If driven constantly by noise, this might generate a symmetric triplet with fuzzy side-peaks.  This is of course rather speculative as it is not known if such phase locked states exist nor how much excitation could be achieved by the turbulence noise.

\subsection{Long Period Blazhko}
There are also cases with very long periods, up to 10 years \citep[see e.g. Figure~5 in][]{Soszynski}, which seem to be at odds with the idea that nonlinear coupling should cause phase-locking for modes that are close together.  One explanation is that the nonlinear coupling between the modes could be too weak for phase locking to occur.  The coupling coefficient depends on the spacial and temporal characteristics of the modes and the order of the resonance involved and so it can vary substantially depending on the specific modes involved.  There can also be an element of chance involved:  If the coefficient is expected to lie in a range from negative to positive values then for certain stellar parameters it may ``accidentally" have a value very near zero.

Another explanation is that these stars could have extremely low rotation rates.  In this case, a dipolar nonradial mode could have its axis drift due to a combination of random forces from turbulence plus the effects of the slow rotation.  If this mode was phase-locked to the fundamental, then this could be described as a noisy version of the \citet{Nowakowski} model (see discussion in Section \ref{introduction}) and the observed Blazhko frequency would be related to (but not equal to) the rotational frequency.  As noted before, this model is also a beating-modes model since the amplitudes of the modes are not actually being modulated.

This long period case is also problematic for Blazhko models of the ``unstable resonance" type such as the ninth overtone model (Section~\ref{ninthmodel}).  One might expect the system in that case simply go to stable phase-locking of the two modes rather than oscillate with an exceedingly long period---probably far longer than characteristic time constants of the system, such as the damping time constants for the modes involved.  The time scale is also excessively long for the shock model (Section~\ref{shockmodel}).  It might be compatible with the Stothers model which seems to be more viable for very long period modulation (Section~\ref{stothersmodel}).

\section{PROBLEMS WITH OTHER BLAZHKO EXPLANATIONS}
\label{problems}
\subsection{The Ninth Overtone Model}
\label{ninthmodel}
One explanation for the Blazhko effect (arguably the leading one) involves an unstable resonant interaction between the fundamental mode and the ninth radial overtone \citep{Szabo, Buchler2011, Kollath2011}.  This is a very high order resonance with the frequency ratio of the overtone to the fundamental being 9 to 2.  As a result, correspondingly high powers appear in the amplitude equations \citep[Equation 2 in][]{Buchler2011}.  This can cause extreme changes in coupling strength in response to moderate changes in amplitude.

Probably the most serious problem is that the spectral peak that they identify with the overtone is exceedingly small compared to the fundamental.  Resonance based models like theirs depend on a nonlinear interaction between the modes to generate the Blazhko oscillation.  This interaction must exhibit an instability that results in a substantial amount of energy being transferred back and forth at the Blazhko frequency between the fundamental and the other mode or modes. It is this energy shift that causes the observed amplitude changes in the fundamental during the Blazhko cycle.  When the energy is shifted away from the fundamental it goes into the ninth overtone.  One can see this quite clearly in the mode simulations of \citet{Buchler2011}.  In their Figure~1, note that the amplitude curve of mode B is roughly an inverted copy of mode A and is comparable in size.  Also \cite{Buchler1984} state in their introduction ``... one can have periodic or `chaotic' energy transfer between the resonant modes."  But this appears to be inconsistent with the negligibly small size of the ninth overtone peak in the observational spectrum.  To be specific, from \citet{Szabo} Table~1 and Figure~8, the amplitude of the peak at $f_0$ (the fundamental) is about 158 mmag and the peak at $9/2f_0$ (the presumed ninth overtone) is about 1.65 mmag, about 100 times smaller.  (The size discrepancy is even worse than it appears, since the kinetic energy in a mode is proportional to the square of the velocity, and the luminosity is known to vary roughly in proportion to velocity, making the energy ratio on the order of 10000.)  When there is a large size discrepancy between the two modes there is a corresponding imbalance in the interaction between them.  This is clear from the amplitude equations of  \citet{Buchler2011} (their Equation 2) which shows that when $B$ is much smaller than $A$, the effect of $B$ on the dynamics of $A$ is negligible, while the effect of $A$ on the dynamics of $B$ is quite strong.  To make an analogy, an ocean wave can have a big effect on the motion of a surfer, while the surfer has a negligible effect on the wave.  Thus the dynamics of the fundamental will proceed as if the overtone did not exist and there can be no Blazhko effect.

A second problem, is that the ninth overtone is being identified as a half integer harmonic of $f_0$, but in addition to the expected peak at $9/2 f_0$ the spectrum also shows peaks at $1/2 f_0$, $3/2 f_0$, $5/2 f_0$, $7/2 f_0$, etc.  Of these, peak at $3/2 f_0$ is the largest by far, and is close to 3 times the amplitude of the one at $9/2 f_0$. This suggests that $3/2 f_0$ is the peak associated with an excited mode, while the others, including the peak at $9/2 f_0$, may be nonlinear mixing product peaks.  The fact that the $9/2 f_0$ peak is slightly elevated compared to the $7/2 f_0$ peak may simply reflect the fact that $9/2 f_0$ is the third harmonic of $3/2 f_0$.

A third problem is that the alternating peak heights in the Kepler time series data appear and disappear in an intermittent, seemingly random fashion, rather than varying in synchronization with the Blazhko cycle.  If the half integer harmonics are directly related to the Blazhko effect then they must play an active role in the Blazhko dynamics.  They cannot be exhibiting their own separate dynamics and still be held responsible for the Blazhko effect.  This is in contrast to the \cite{Smolec} test of the Stothers model, where they state ``We note that in our models period doubling is strictly repetitive and always appears at the same phase of the Blazhko cycle."

A fourth problem is that since the ninth overtone model is entirely radial, the Blazhko effect should be reproducible with the available 1D hydrocodes.  Apparently this has not occurred.

\subsection{The Shock Model}
\label{shockmodel}
The shock model of \cite{Gillet} presents just the framework of a possible theory with all of the needed details either missing or fatally flawed.  The basic idea is that shock waves build up in the atmosphere while the fundamental is growing in amplitude, then these shocks generate turbulence affecting the excitation mechanism for the fundamental, then the fundamental decays in amplitude and finally the atmosphere becomes stable again and the process repeats.  Gillet claims this process depends critically on the presence of the ``transient" first overtone, but this concept is not clearly defined.  Does transient mean the mode is only present for part of the Blazhko cycle?  If so what portion of the cycle and why does Gillet not look at the observational data to try and prove that the overtone has a transient presence?  Gillet admits that the observed amplitude of the first overtone is about 100 times smaller than the fundamental.  From this one might estimate that the energy could be on the order of 10000 times smaller since energy typically is proportional to the square of the velocity while the light signal tends to be roughly proportional to the velocity.  So how can something so weak and insignificant be expected to play a major role in the stellar dynamics?  Gillet also admits that ``The physical origin of shock s3' has not been determined yet" (i.e. it is complete speculation that is has anything to do with the first overtone).  Also there is the fact that the first overtone has a different frequency than the fundamental so if it is supposed to be generating a shock wave at a certain phase of the fundamental pulsation, there will be a problem since the overtone will have a different relative phase each successive pulsation.  The results also depend on the ability of the shock wave to suddenly cause the complete ``desynchronization" of the photospheric layer; is there evidence that a shock wave can in fact do this and do it suddenly rather than gradually?  One should bear in mind that these shock waves are not caused by an outside source, but simply appear as part of the dynamics of the pulsational motion.  So it seems surprising that such a shock wave would suddenly disrupt the motion that is generating it.  It would seem more likely that if turbulence was emerging this would limit the strength of the shock wave and an equilibrium would be reached.  Aside from the turbulence associated with the desynchronization this would seem to be an entirely radial model and so there is also the question as to why the proposed mechanism is not seen in 1D hydrocode simulations.  \citet{Fossati} discuss this model---they find that the changes in the level of turbulence required by the model appears higher than what should be expected from their measurements of variations in the microturbulent velocity for RR Lyr.

\subsection{The Stothers Model}
\label{stothersmodel}
The Stothers model has received a fair amount of attention recently.  It is discussed at length in \citet{Kovacs} and the papers by \citet{Smolec} and \citet{Molnar2012} are entirely devoted to it.  It depends on several assumptions that are very difficult to test.  In particular it is assumed that there is a magnetic field whose amplitude varies at the Blazhko frequency and that the changes in magnetic field in turn affect the level of turbulence sufficiently to modulate the amplitude of the fundamental mode.  In an effort to test the theory, \citet{Smolec} use a hydro model into which the assumed periodic changes in the turbulence are introduced in an ad hoc manner.  They are able to recreate some of the desired Blazhko features including asymmetric side-peaks and the disappearance of the bump during the ascending Blazhko phase, but they think that the amount of turbulence changes required for this may unrealistic.  The simulation also fails to reproduce the upward motion of the bump as seen in Figure~\ref{dam} of the current paper.  The behavior of the bump in the Stothers case is of course a function of the level of turbulence.  Also there is no significant amplitude differential between the upper and lower Blazhko envelope functions.  \citet{Molnar2012} using a different hydro model find the Stothers model could be workable but only for Blazhko periods longer than 100 days due to the slow response of the system.  \citet{Fossati} discuss this model---they find that the changes in the level of turbulence required by the model appears higher than what should be expected from their measurements of variations in the microturbulent velocity for RR Lyr.

\section{RELATION TO EARLIER WORK BY THE AUTHOR}
\label{relation}
In a previous paper \citep{Bryant} it was shown that an accurate fit to second quarter \emph{Kepler} data for RR Lyr could be achieved by a hybrid mode consisting of two component modes of the same frequency.  The first of these components is a highly non-sinusoidal sawtooth-like waveform, while the second is approximated as a sinusoid.  Although locked in frequency, the two components are allowed to change independently as a function of the Blazhko phase.  The first component varies little in amplitude or frequency throughout the Blazhko cycle, while the second changes quite strongly both in amplitude and in phase relative to the first component.  By optimizing the fit to the \emph{Kepler} data, the amplitude and phase behavior was obtained and displayed in Figure~4 of \citet{Bryant}.

The current work is related to that earlier work in the following way:  It is found that the phase and amplitude dynamics of the second component of the earlier description can be approximately generated by adding together two constant amplitude sine waves, one with frequency $\omega_1$ and a second of lower amplitude with frequency $\omega_1 + \omega_B$.  This was noted previously by \citet{Guggenberger2008} who found this type of phase and amplitude behavior in several RR Lyrae stars and attributed it to this type of beating process.  The first component corresponds to PO1a in the current model.  The second sine wave corresponds to PO2.  The first sine wave corresponds to PO1b, although in the current model this is made non-sinusoidal.  Making this non-sinusoidal helps to correct one of the problems with that earlier model, in that the bump feature generated by that model was not as pronounced as in the \emph{Kepler} data.

\section{CONCLUSION}
It has been shown that a simple beating-modes model of the Blazhko effect can provide a remarkably good fit to \emph{Kepler} data for RR Lyr.  The model consists of two periodic oscillations of slightly different frequency that are simply added together.  This is in contrast to the much more complex picture of resonant mode models that depend on nonlinear coupling between modes leading to an oscillatory exchange of energy between them.  The current model reproduces for the first time the apparent motion, disappearance and reappearance of the bump feature, the phase difference between the upper and lower modulation envelope functions and their relationship to the motion of the bump.  In addition, it has been shown that the amplitude disparity between the upper and lower envelope functions can be reproduced if the light curve is assumed to be a nonlinear function of the radial velocity.  Previous work by \citet{Cox1993,Cox2009} suggests that the modes involved could be the fundamental and the $g_4$, $l=1$ nonradial mode.  The model is easily extended to cases exhibiting multiple modulation frequencies by simply associating additional large near-resonant peaks in the spectrum with active modes of the star, i.e. one additional mode is required for each additional modulation frequency.  Application of this extended model was discussed for the case of V445 Lyr.  The mysterious double-maxima waveform observed for that case is explained providing \emph{very strong evidence} that the hypothesis of this paper is correct.

In addition, problems with other recent models of the Blazhko effect have been discussed in some detail, including the ninth overtone model, the shock model and the Stothers model.

\acknowledgements
The author thanks Katrien Kolenberg for very helpful discussion and comments about the model, the manuscript and the spectral analysis.  The author thanks Elisabeth Guggenberger for helpful comments on the manuscript.

\appendix
\section{ONE-ZONE MODEL AND PO1 DETAILS}
\label{one-zone}
The equations of motion are derived from a simplified analysis of pressure and gravity acting on the portion of a star above a certain radius $R_0$.  We may take $R_0$ to be the effective bottom of the stellar envelope if we are considering the fundamental mode (as we are in this paper) or the outermost radial node for other modes such as the first radial overtone.  For large amplitude oscillations, the motion is ballistic or gravity dominant for most of the cycle, with pressure becoming dominant only near the minimum position of the cycle.  This leads to a ``bouncing ball" type of oscillation whose derivative is the sawtooth-like waveform mentioned previously.  Regardless of the accuracy of this simplified picture of the dynamics, the ability to reproduce the sawtooth-like waveform shows that the one-zone model seems to capture an essential aspect of the true dynamics.

The model takes the form of an ordinary differential equation for a harmonic oscillator, but with the usual linear forcing term replaced with a nonlinear function $f(x)$ that, in a very simplified way, captures the essence of the gravity and pressure forces, i.e. $\dot x = v$ and $\dot v = f(x)$ where $x$ and $v$ are position and velocity.  This model is conservative, excitation and damping of the oscillations are omitted, but could be added to the model if desired.  Also currently assumed to be small and ignored are nonlinear interactions between modes that could lead to energy transfer or phase shift.  In \citet{Bryant} the following expression is derived for the forcing function:
\begin{equation}\label{over_force}
f(x) = A(R_1+x)^2/((R_1+x)^3-1)^\gamma - B/(R_1+x)^2,
\end{equation}
where:
\begin{equation}\label{over_A}
A = \omega ^2 V_1 ^\gamma (3 \gamma R_1 ^4 V_1 ^{-1} -4R_1)^{-1},
\end{equation}
\begin{equation}\label{over_B}
B = \omega ^2 R_1 ^4 (3 \gamma R_1 ^4 V_1 ^{-1} -4R_1)^{-1},
\end{equation}
$\gamma = 1+1/n_p$, $V_1 = R_1 ^3 -1$, $R_1$ is a characteristic radius for the stellar material that lies above $r = R_0$ and $n_p$ is the polytropic index.  The problem has been rescaled so that $R_0 = 1$.  To keep the values near unity, time is in units of 0.1 days.  The values used in the model are $\omega = 1.5$ rad/0.1 day, $n_p = 1.5$, $R_1=1.1$, $\omega_1=1.108328$ and $\omega_B= 0.015393$.  The Blazhko period is 72 pulsation periods.  Initial condition for PO1a: $x_{1a}(0)=-0.0691$, $v_{1a}(0)=0$.  Initial condition for PO1b is chosen so that it produces the identical oscillation except that it is leading in phase by 2.1058232 radians or 0.19 days:  $x_{1b}(0)=0.185258$, $v_{1b}(0)=0.078770$.  Note that $\omega$ is the frequency for small amplitude oscillations;  the actual oscillation frequency drops as the amplitude increases and is equal to $\omega_1$ for the specified initial conditions.
The velocity for PO1, $v_1(t)$, is the combined velocity of PO1a and PO1b, with scale factors chosen to fit the \emph{Kepler} data:
\begin{equation}\label{eqnpo1}
v_1(t) =  v_{1a}(t) + 0.15v_{1b}(t)
\end{equation}
This function is plotted in Figure~\ref{po1}.

\section{FREQUENCY SPECTRA FOR QUASIPERIODICITY}
\label{quasi}
When a nonlinear dynamical system is quasiperiodic, it will have two base frequencies which are incommensurate, i.e. they do not have a rational ratio.  In the phase space the dynamics will lie on a two-torus or doughnut surface.  The dynamics will wind its way around this surface, taking an infinite amount of time to completely cover it.  Motion on a 2-torus can be described by two angle-like variables, $\theta_1$ and $\theta_2$, corresponding to the two ways in which one can go around the torus.  We can define the angles so that they satisfy $\theta_1 = (\omega_1 t) \bmod (2\pi)$ and $\theta_2 = (\omega_2 t) \bmod (2\pi)$.  The state of the system can be expressed as a function of these two angles:
\begin{equation}\label{state}
\textbf{\textit{X}} = \textbf{\textit{X}}(\theta_1,\theta_2) = \mbox{state vector of the system}
\end{equation}
The state is $2\pi$ periodic in both angles (since a $2\pi$ shift in either angle corresponds to one pass around the torus).
Thus it can be expanded as a two dimensional Fourier series:
\begin{equation}\label{expan}
\textbf{\textit{X}} = \sum_{m=-\infty}^{\infty} \sum_{n=-\infty}^{\infty} \textbf{\textit{A}}_{mn} \exp(im\theta_1+in\theta_2)
\end{equation}
where
\begin{equation}\label{amps}
\textbf{\textit{A}}_{mn} ={1\over (2\pi)^2}\int_0^{2\pi}\mathrm{d}\theta_1 \int_0^{2\pi}\mathrm{d}\theta_2 \textbf{\textit{X}}(\theta_1,\theta_2)\exp(-im\theta_1-in\theta_2)
\end{equation}
In terms of time this becomes:
\begin{equation}\label{time_expan}
\textbf{\textit{X}} = \sum_{m=-\infty}^{\infty} \sum_{n=-\infty}^{\infty} \textbf{\textit{A}}_{mn} \exp(i \omega_{nm}t)
\end{equation}
where the frequencies in the spectrum are
\begin{equation}\label{freqs}
\omega_{mn} = m\omega_1 + n\omega_2
\end{equation}
The amplitudes $\textbf{\textit{A}}_{mn}$ can also be obtained as an integral over time:
\begin{equation}\label{time_amps}
\textbf{\textit{A}}_{mn} = \lim_{T \to \infty}{1\over T}\int_0^T\mathrm{d}t \textbf{\textit{X}}(t)\exp(-i\omega_{mn}t)
\end{equation}
The amplitude of these components must fall off as $m$ and $n$ become large.

Cases of three frequency quasiperiodicity (or higher) require a simple extension of the above analysis to a higher dimensional Fourier series.

\end{document}